\documentclass[12pt,preprint]{aastex}

\begin{document}


\title{Detection of a $60\arcdeg$-Long Dwarf Galaxy Debris Stream}

\author{C. J. Grillmair}
\affil{Spitzer Science Center, 1200 E. California Blvd., Pasadena,  CA 91125}
\email{carl@ipac.caltech.edu}

\begin{abstract}

We report on a $60\arcdeg$-long stream of stars, extending from Ursa
Major to Sextans, in the Sloan Digital Sky Survey. The stream is
approximately $2\arcdeg$ wide and is clearly distinct from the
northern tidal arm of the Sagittarius dwarf galaxy. The apparent width
of the stream indicates a progenitor with a size and mass similar to
that of a dwarf galaxy. The stream is about 21 kpc distant and appears
to be oriented almost perpendicular to our line of sight. The visible
portion of the stream does not pass near any known dwarf galaxies,
though we cannot rule out that the stream may form the inner part of a
known dwarf galaxy's orbit. The most likely explanation is that the
stream constitutes the remains a dwarf galaxy that has been completely
disrupted at some point in the past. We also briefly report on the
discovery of a diminutive Galactic satellite which lies near the
projected path of the new stream, but is unlikely to be related to it.

\end{abstract}



\section{Introduction}
The Sloan Digital Sky Survey (SDSS) continues to be a remarkable
resource for studies of Galactic structure and the Local Group. In
addition to the large scale features attributed to past galaxy
accretion events \citep{yann03,maje2003,roch04}, SDSS data was used to
detect the remarkably strong tidal tails of Palomar 5
\citep{oden2001,rock2002, oden2003, grill2006b}. Most recently,
\citet{will2005}, \citet{zuck2006}, and \citet{belo2006b} used SDSS
data to discover several new dwarf satellites of the Milky Way.

Substantial tidal streams have now been found associated with at least
two of the eight globular clusters in the SDSS area; Pal 5
\citep{oden2003, grill2006b} and NGC 5466 \citep{belo05,
grill2006a}. Most recently, \citet{grill2006c} discovered a
$63\arcdeg$ cold stellar stream without an obvious progenitor. Though
the parent body may be one of the $\approx 140$ globular clusters
situated outside the SDSS area, it is equally possible that the stream
is all that remains of a globular cluster that was disrupted long ago.

In this paper we analyze a broad, $60\arcdeg$ long stream,
independently detected in SDSS data by \citet{belo2006b}. We briefly
describe our analysis in Section \ref{analysis}. We discuss the
detection of the stellar stream in Section \ref{discussion}, make
some initial distance estimates in Section \ref{distance}, and attempt
to identify a progenitor in Section \ref{progenitor}. We make
concluding remarks in Section \ref{conclusions}.

\section{Data Analysis \label{analysis}}

Data comprising $ugriz$ photometry for $5.3 \times
10^7$ stars in the region $124\arcdeg < \alpha < 251\arcdeg$ and
$-1\arcdeg < \delta < 65\arcdeg$ were extracted from the SDSS database
using the SDSS CasJobs query system. The data were analyzed using the
matched filter technique employed by \citet{grill2006a},
\citet{grill2006b}, and \citet{grill2006c}, and described in detail by
\citet{rock2002}. This technique is made necessary by the fact that,
over the magnitude range and over the region of sky we are
considering, the foreground disk stars outnumber the more distant
stars in the Galactic halo by three orders of magnitude. Applied in
the color-magnitude domain, the matched filter is a means by which we
can optimally differentiate between two populations, provided we know
the color-magnitude distribution of each.

We used the SDSS photometry to create a color-magnitude density or
Hess diagram for both stars of interest and for the contaminating
foreground population. Dividing the former by the latter, we generated
an array of relative weights which constitutes an optimal
color-magnitude filter.  Using this filter, every star in the survey
can be assigned a weight or probability of association with a
particular color-magnitude sequence.  Having used observed data to
generate it, the filter implicitly includes the effects of photometric
uncertainties.

To generate the filters, we used the SDSS-observed color magnitude
distributions for stars in each of the eight globular clusters in the
SDSS DR4 area (NGC 2419, Pal 3, NGC 5272, NGC 5466, Pal 5, NGC 6205,
NGC 7078, and NGC 7089).  A single Hess diagram for field stars was
generated using $1.2 \times 10^7$ stars spread over $\sim 2200$
deg$^2$ of DR4.  We applied each of the eight optimal filters in turn
to the entire survey area, and the weighted star counts were summed by
location on the sky to produce two dimensional probability maps.

We used all stars with $15 < g < 22.5$. We dereddended the SDSS
photometry as a function of position on the sky using the DIRBE/IRAS
dust maps of \citet{schleg98}. We optimally filtered the $g - u$,
$g - r$, $g - i$, and $g - z$ star counts independently and then
co-added the resulting weight images.  In Figure 1 we show the final,
combined, filtered star count distribution, using a filter matched to
the color magnitude distribution of stars in M 13. The image has been
smoothed with a Gaussian kernel with $\sigma = 0.2\arcdeg$. A
low-order, polynomial surface has been subtracted from the image to
approximately remove large scale gradients due to the Galactic disk
and bulge.

\section{Discussion \label{discussion}}

Quite obvious in Figure 1 are two broad streams, one running east to
west across the center of the field, and the other running more nearly
north and south. The stronger, east-west stream is part of the well
known northern tidal tail of the Sagittarius dwarf galaxy
\citep{maje2003,mart2004}. The north-south stream is clearly distinct
from the Sagittarius stream (intersecting at an angle of almost
$45\arcdeg$ on the sky), and runs from (R.A., decl) = ($163\arcdeg,
-1\arcdeg$) to approximately (R.A., decl.) = ($141\arcdeg,
52\arcdeg$). On the sky, the stream runs in a $60\arcdeg$, nearly
great circle path from Ursa Major in the north to Sextans in the
south. The stream is distinct from the low latitude halo stream found
by Yanny et al. (2003), which lies some $30\arcdeg$ to the east of the
present stream. The stream is also visible in Figure 1 of \citet{belo2006b}.

The stream was found using a filter that is matched to the color
magnitude distribution and luminosity function of stars in the
globular cluster M 13, though shifted faintwards by 2.1 magnitudes. We
used M 13's color-magnitude locus primarily because, owing to 13's
relative proximity, we have a better measure of its luminosity
function and the effects of SDSS completeness.

The stream is not a product of our dereddening procedure. Figure 2
shows the portion of the reddening map of \citet{schleg98} covering
the region of interest. There appears to be no correlation between the
new stream and the applied reddening corrections. The maximum values
of $E(B-V)$ are $\approx 0.05$, with typical values of 0.03 in the
southern region of the stream and 0.01 in the northern.  There are
diminutions here and there in the stream that could perhaps be
attributed to regions of higher reddening in Figure 2, but there is
certainly no long trough in the reddening map that could be held to
account for the entire stream.

We have also run our optimum filter against the SDSS DR4 galaxy
catalog to investigate whether the stream could be due to confusion
with faint galaxies. Though there are occassional concentrations of
galaxies along the stream's trajectory, there is no indication of a
continuous north-south feature as is apparent in Figure 1. We conclude
that faint galaxies also cannot be held to account for the stream.

At its southern end the stream is truncated by the limits of the
available data. The apparent surface density of the stream is highest
at this point, and we are fairly confident that larger surveys will be
able to extend the current stream substantially to the south. On the
northern end, the stream becomes indiscernible beyond $\delta =
51\arcdeg$. We have attempted to trace the stream in the portion of
the SDSS extending from $52\arcdeg < \delta < 65\arcdeg$ by shifting
the filter to both brighter and fainter magnitudes, but to no
avail. 

We conclude that we are seeing either $(1)$ a substantial local drop in
volume density at the northern end of the stream (due perhaps to an
episodic stripping process or a higher velocity portion of an
eccentric orbit), $(2)$ a significant change in the color-magnitude
distribution of stars along the stream, or $(3)$ the physical end of
the stream. We consider the second option to be the least likely of the
three possibilities, though it is not inconceivable that distinct
stellar populations could have been pulled from the progenitor in
sequence. Arguments for or against $(i)$ and ($iii)$ will have to
await the acquisition of survey data covering a larger field.

Sampling by eye at several representative points, the stream appears
to be about $2\arcdeg$ wide on average. This is significantly broader
than the globular cluster streams found by \citet{oden2003,
grill2006a, grill2006b} and \citet{grill2006c}. On the other hand, it
is similar to the western portion of the Sagittarius stream visible in
Figure 1. $2\arcdeg$ corresponds to about 700 pc at our estimated
distance to the stream (see below). If the stream is circular in cross
section, then in a logarithmic potential with $v_c = 220$ km s$^{-1}$,
a radial width of this magnitude would correspond to $\Delta E > 20$
km s$^{-1}$. This is considerably larger than the expected random
velocities of stars weakly stripped from globular clusters and implies
a much larger progenitor mass. Of course, with the stream passing high
over the outer regions of the disk ($165\arcdeg < {\it l} <
252\arcdeg, 44\arcdeg < b < 50\arcdeg$) we cannot currently say how
broad the stream is along a radial from the Galactic center. However,
700 pc remains much larger than the tidal diameters of globular
clusters, and we conclude that the progenitor is most likely to be an
extant or disrupted dwarf galaxy.

Integrating the background subtracted, weighted star counts along the
stream over a width of $\approx 2\arcdeg$ we find that the total
number of stars in the discernible stream down to $g = 22.5$ is $1250
\pm 200$. The surface density of stars in the southern portion of the
stream ($-1\arcdeg < \delta < 14\arcdeg$) is roughly twice that of the
northernmost section ($39\arcdeg < \delta < 52\arcdeg$). The highest
surface densities along the southern portion of the stream are $\sim
40$ stars deg$^{-2}$.

\subsection{Distance to the Stream \label{distance}}

The power of the matched filter resides primarily at the main sequence
turn-off and below, where the stellar luminosity function increases
rapidly and the bulk of the foreground contaminants lie well to the
red. If our filter is globally matching the color-magnitude
distribution of the stream population, then we can use the filter
response to estimate distances by main sequence fitting
(e.g. \citet{grill2006c}).  In Figure 3 we show the color-magnitude
distribution for the stream stars, extracted by generating a Hess
diagram of stars in a 11 deg.$^2$ area covering the southern portion
of the stream, and subtracting a similar field star distribution
sampled over 28 deg$^2$ to the east and west of the stream.  A clear
signature of the sub-giant branch and the main sequence turn-off is
evident in the stream population.  Moreover, the sub-giant and
turn-off region of the distribution match the dereddened, shifted $g,
g - i$ RGB/main sequence locus of M 13 very well.

Varying the magnitude shift applied to M 13's main sequence locus from
1.5 to 3.0 mags, we measured the foreground-subtracted, mean surface
density of stream stars in the regions $-1\arcdeg < \delta <
14\arcdeg$, $29\arcdeg < \delta < 39\arcdeg$, and $39\arcdeg < \delta
< 52\arcdeg$.  To avoid potential problems related to a difference in
age between M 13 and the stream stars, we used only the portion of the
filter with $19.5 < g < 22.5$, where the bright cutoff is 0.8
magnitudes below M 13's main sequence turn-off. This reduces the
stream contrast significantly, but still provides sufficient
signal to identify the optimum magnitude shift.

Fitting Gaussians to the mean surface densities as a function of
magnitude shift (e.g. \citet{grill2006c}, we find the highest
contrasts occur for shifts of 2.1, 2.4, and $2.15 \pm 0.1$ mags for
the three stream segments identified above.  This puts the stream
roughly perpendicular to our line of sight at an average distance 2.78
times that of M 13. Adopting a distance to M 13 of 7.7 kpc, we arrive
at a mean Sun-stream distance of $21.4 \pm 1$ kpc, where the
uncertainty reflects only random measurement errors.

The measurement of relative distances in this manner rests on the
assumption that the color-magnitude distribution of stream stars is
uniform, and that different parts of the stream will respond to
magnitude shifts in the filter in the same way. Of course, this may
not be true if different sub-populations of the parent galaxy were
removed at different times.  Our relative distance estimates may also
be subject to variations in SDSS sensitivity and completeness at faint
magnitudes, though over the scales with which we are dealing, it seems
reasonable to suppose that the such variations will have largely
averaged out. 

\subsection{The Stream Progenitor \label{progenitor}}

The stream does not pass near any of the known dwarf galaxies in the
Local Group. While Leo I and Leo A lie close to the stream in
projection, their respective distances of $254 \pm 19$ kpc
\citep{bell2004} and $800 \pm 40$ kpc \citep{dolph2002} put them well
beyond the 21 kpc distance we determined above.  The Sextans dwarf is
situated about $7\arcdeg$ west of the southern end of the
stream. However, at its distance of $86 \pm 6$ kpc \citep{mateo95} it
too is too far away to be a likely progenitor. At this
distance its main sequence turn-off would be well below the
completeness limit of the SDSS data, which is clearly inconsistent
with Figure 3.

Though we are obviously limited by the lack of velocity information,
for a given model of the Galactic potential the progenitor's orbit is
actually fairly well constrained by the observed distance and
orientation of the stream. Using the model of \citet{allen91} (which
includes a disk, bulge, and spherical halo, and which
\citet{grill2006a} and \citet{grill2006b} found to work reasonably
well for NGC 5466 and Pal 5), we use a least squares method to fit
both the orientation on the sky and the distance measurements in
Section \ref{distance}. In addition to a number of normal points lying
along the centerline of the stream, we chose as a velocity fiducial
point a position at the northern tip of the stream at (R.A., decl) =
(141.2\arcdeg, 51.2\arcdeg).

If we allow the proper motions to be free ranging and uninteresting
parameters, we find that the 95\% confidence interval in $\chi^2$
predicts $-35 < v_{LSR} < +57$ km s$^{-1}$ at the fiducial point.
The corresponding ranges in perigalactic and apogalactic radii are
$9.1 < R_p < 9.3$ kpc and $28 < R_a < 32$ kpc. Of course, these ranges
do not take into account uncertainties in the absolute distance of the
stream (which depends on the uncertainty in M 13's distance) or in the
validity of \citet{allen91}'s Galactic model. However, they do suggest
a fairly compact orbit with moderate eccentricity.

Integrating orbits for parameter sets spanning the range above, we
find that there are another five known dwarf galaxies (NGC 185, the
Aquarius dwarf galaxy, NGC 6822, and the Sagittarius irregular and
dwarf elliptical galaxies) that lie within $10\arcdeg$ of the
projected orbits. However, with the exception of the Sagittarius dE,
their measured distances are all greater than 500 kpc and would
therefor require rather unreasonably high orbital eccentricities
($\epsilon > 0.98$). We attribute the proximity of the Sagittarius dE
to the new stream's projected orbit to the expected confluence of
orbit projections in the direction of the Galactic center, and not to
any physical association between them.

In the upper right-hand corner of Figure 1 (R.A., decl.) =
($132.857\arcdeg, +63.136\arcdeg$) is a fairly pronounced
concentration of stars about 37 kpc distant which we identify as a new
globular cluster or dwarf galaxy \citep{grill2006d}.  The object lies
within $5\arcdeg$ of the projected orbit of the stream (1.8 kpc at the
distance of the stream). However, its diminutive size and its 16 kpc
separation from the stream's orbit argue against its being either the
progenitor or a part of the debris in the stream.

\section{Conclusions \label{conclusions}}

Applying optimal contrast filtering techniques to SDSS data, we have
detected a broad stream of stars some $60\arcdeg$ long on the sky. We
are at present unable to identify a progenitor for this stream, though
from its appearance and location on the sky, we believe it to be a
either an extant or disrupted dwarf galaxy. The color-magnitude
distribution of stars in the stream closely matches that of the
globular cluster M 13, indicating that the stars making up the stream
are old and metal poor.  The stream appears to be about 21 kpc
distant, and the portion of the stream contained in the survey data is
roughly perpendicular to our line of sight. 

Refinement of the stream's orbit will require radial velocity
measurements of individual stars along its length.  Ultimately, the
vetted stream stars will become prime targets for the Space
Interferometry Mission, whose proper motion measurements will enable
very much stronger constraints to be placed on both the orbit of the
progenitor and on the potential field of the Galaxy.

\acknowledgments

Funding for the creation and distribution of the SDSS Archive has been
provided by the Alfred P. Sloan Foundation, the Participating
Institutions, the National Aeronautics and Space Administration, the
National Science Foundation, the U.S. Department of Energy, the
Japanese Monbukagakusho, and the Max Planck Society.

{\it Facilities:} \facility{Sloan}.

\clearpage



\begin{figure}
\epsscale{0.8}
\plotone{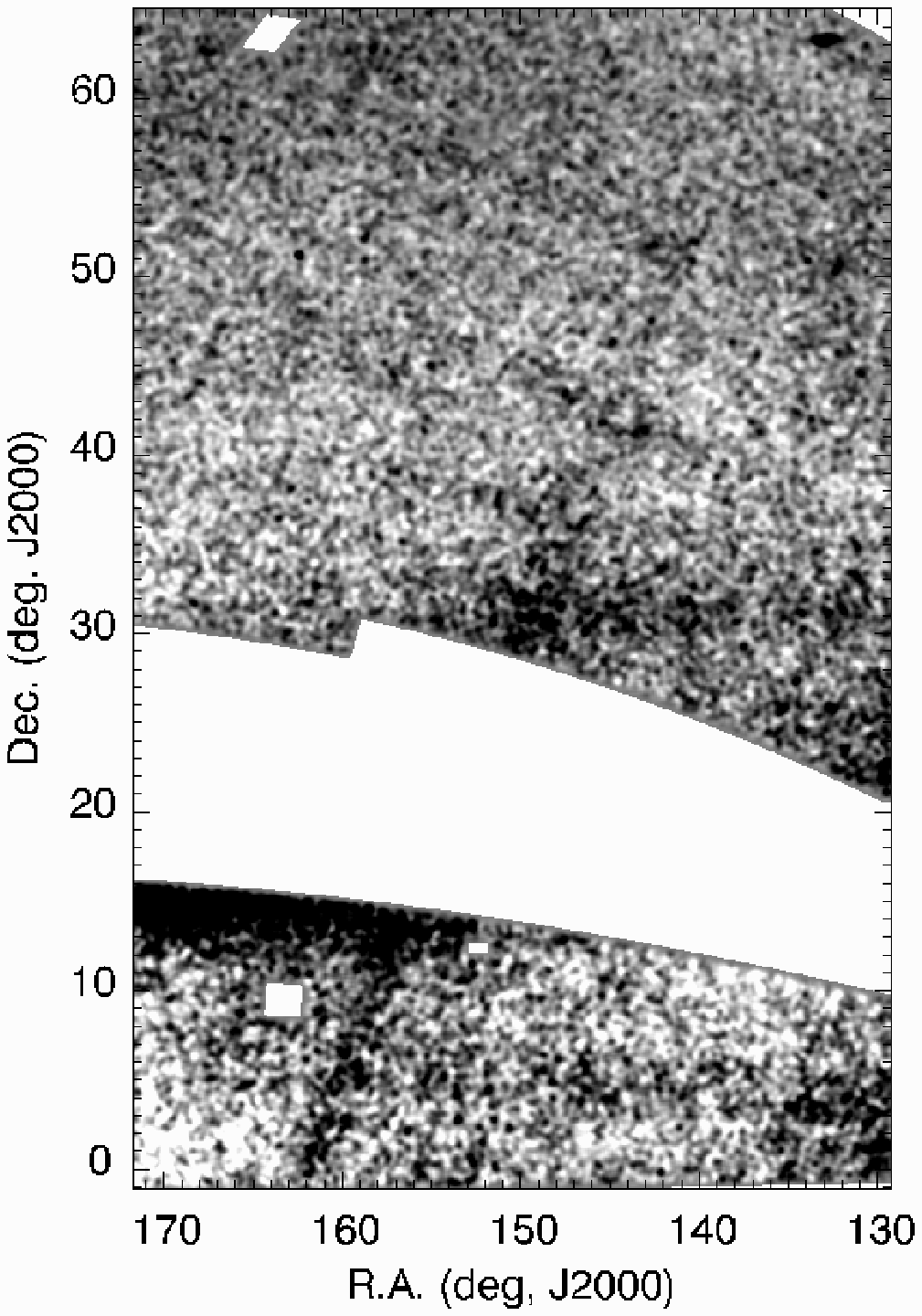}
\caption{Smoothed, summed weight image of the SDSS field after
subtraction of a low-order polynomial surface fit. Darker areas
indicate higher surface densities. The weight image has been smoothed
with a Gaussian kernel with $\sigma = 0.2\arcdeg$. The white areas are
either data missing from DR4, or clusters or bright stars which have
been masked out prior to analysis. The northern tidal arm of the
Sagittarius dwarf galaxy runs roughly east-west above and below the
gap in DR4. The new stream runs from (R.A., decl) = ($163\arcdeg,
-1\arcdeg$) to approximately (R.A., decl.) = ($141\arcdeg,
52\arcdeg$). At the extreme upper right is a new Milky Way satellite
which is considered in detail by \citet{grill2006d}. \label{fig1}}
\end{figure}

\begin{figure}
\epsscale{0.8}
\plotone{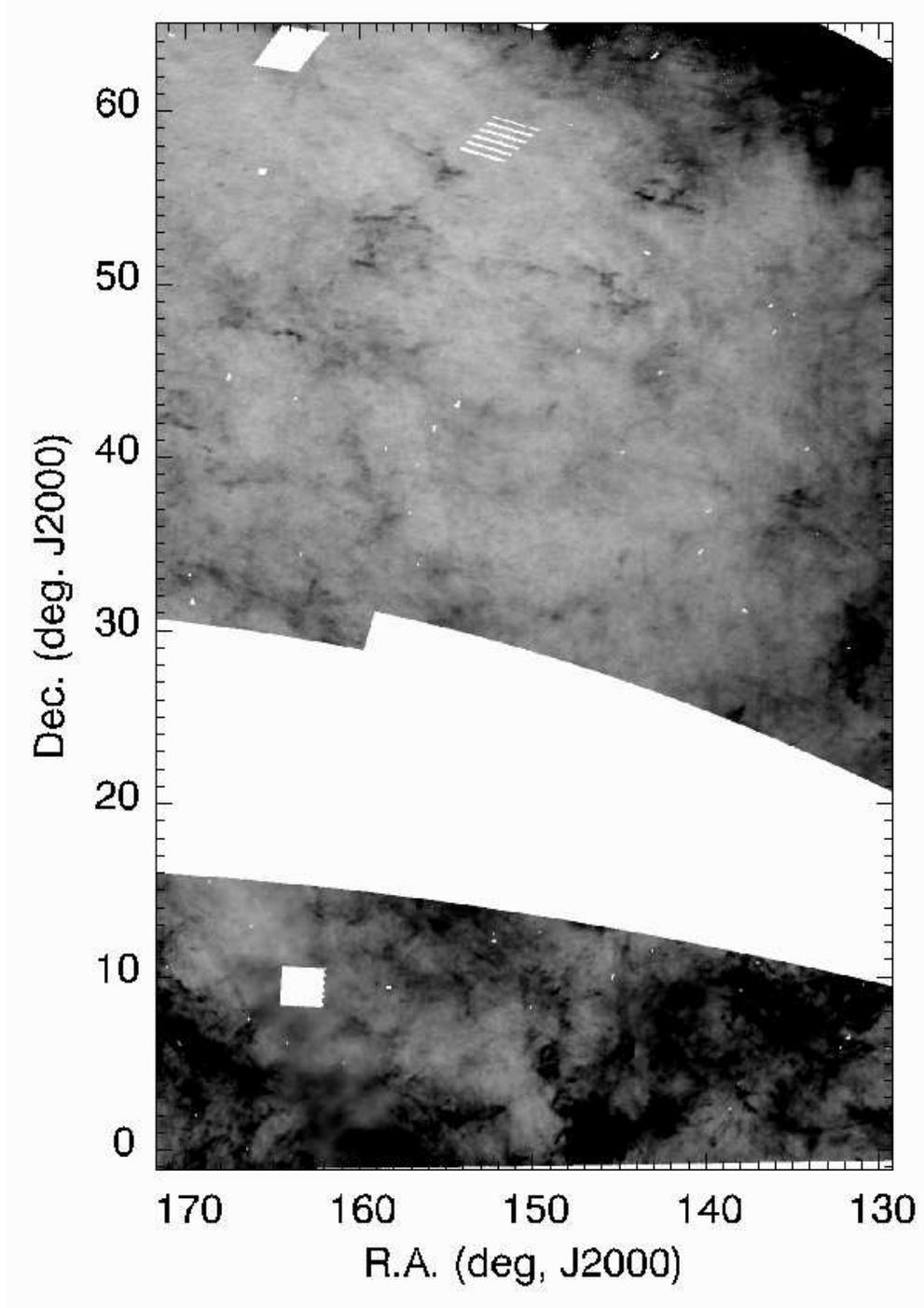}
\caption{Distribution of E(B-V) over the field shown in Figure 1.
Darker areas indicate higher color excesses. Typical values of the color
excess range from 0.01 along the northern part of the new stream to
0.03 along the southern part. \label{fig2}}
\end{figure}

\begin{figure}
\epsscale{0.6}
\plotone{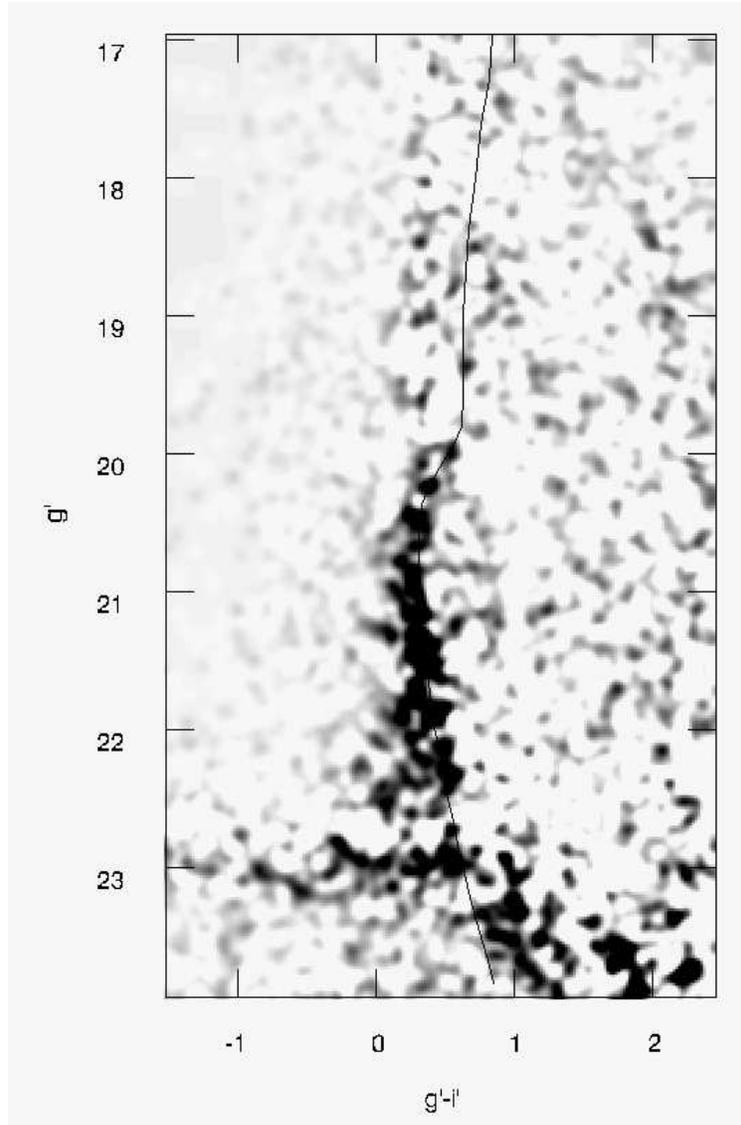}
\caption{The dereddened, background subtracted, color-magnitude
distribution of stream stars. The stream stars include all stars
situated in an 11 deg$^2$ area covering the southern portion of the
stream in Figure 1. The distribution of field stars was computed from
all stars in a 28 deg$^2$ area bracketing the stream to the east and
west. A subgiant branch and main sequence turn-off are clearly
distinguishable. The solid line shows the dereddened locus of giant
branch and main sequence stars as measured in DR4 for M 13, 
shifted faintwards by 2.1 mags.}
\end{figure}

\end{document}